\documentclass[preprint,12pt]{elsarticle}
\usepackage{amssymb}
\usepackage{epsfig,graphicx,bm}
\journal{Icarus}
\begin{document}
\begin{frontmatter}
\title{Surface ages of mid-size Saturnian satellites}
\author[1,2]{Romina P. Di Sisto}
\ead{romina@fcaglp.unlp.edu.ar}
\author[2]{Macarena Zanardi}
\address[1]{Facultad de Ciencias Astron\'omicas y Geof\'\i sicas, Universidad Nacional de La Plata}
\address[2]{Instituto de Astrof\'{\i}sica de La Plata, CCT La Plata-CONICET-UNLP,  Paseo del Bosque 
S/N (1900), La Plata, Argentina}
\begin{abstract}
The observations  of the surfaces of the  mid-sized Saturnian satellites made by Cassini-Huygens mission  have shown a variety of features that allows study of the  processes that took place and are taking place on those worlds. 
Research of the Saturnian satellite surfaces has clear implications not only for Saturn's history and Saturn's surroundings, but also for the Solar System.
Crater counting from high definition images is very important and could serve for the determination of the age of the surfaces.  In a recent paper, we have calculated the production of craters on the mid-sized Saturnian satellites  by Centaur objects  considering the current configuration of the Solar System. Also, we have compared our results with crater counts from Cassini images by other authors and we have noted that  the number of observed small craters is less than our calculated theoretical number. 
In this paper we estimate the age of the surface for each observed terrain on each mid-sized satellite of Saturn. All the surfaces analyzed appear to be old with the exception of Enceladus. However, we have noticed that  since there are less observed small craters than calculated (except on Iapetus), this results in younger ages than expected. This could be the result of efficient endogenous or exogenous process(es)  for erasing small craters and/or crater saturation at those sizes.  The size limit from which the observed number  of smaller craters is less than the calculated is different for each satellite, possibly indicating processes that are unique to each, but other potential common explanations for  this paucity of small craters would be crater saturation and/or deposition of E-ring particles.
These processes are also suggested by the findings that the smaller craters are being preferentially removed, and  the erasure process is gradual.

 On Enceladus, only mid and high latitude plains have remnants of old terrains; the other regions could be young. 
In particular, the regions near the South Polar Terrain could be as young as 50 Myrs old.
On the contrary for Iapetus, all the surface is old and it notably registers  a primordial  source of craters. As the crater size is decreased, it would be perceived to approach  saturation until $D \lesssim 2$ km-craters, where saturation is complete.

\end{abstract}
\begin{keyword}
Saturn, satellites; Cratering; Centaurs
\end{keyword}
\end{frontmatter}

\section{Introduction} 
\label{intro}

The surfaces of mid-sized Saturnian satellites are a kind of laboratory where it is possible to observe and investigate the physical and dynamic processes that have been taking place around Saturn and also throughout the Solar System.  The study of impact craters on the satellite surfaces allows us to better understand what could be happening  in the Saturn environment to produce these craters. 
The Saturnian satellite system was observed and studied in the past by Pioneer 11 and  Voyager 1 and 2 spacecrafts, greatly increasing the knowledge of the Saturn system. Voyager images revealed diverse satellite surfaces (Smith et al., 1981, 1982). Crater counts on those images indicated that Saturnian satellites have been variably cratered, which suggest different geologic histories (Plescia and Boyce 1982, 1983, 1985).        
At present, the  Cassini-Huygens mission is visiting the Saturn system, and the detailed observations it renders provide us with new paradigms and physical processes to understand and interpret. 

The mid-sized icy satellites of Saturn are Mimas, Enceladus, Tethys, Dione, Rhea, and Iapetus. They  
are regular satellites, mainly composed of water ice and in synchronous rotation. The Cassini-Huygens mission has observed all of them in detail allowing scientists to obtain accurate information on the shapes, mean radii and densities (Thomas, 2010) and gravity fields (Jacobson et al., 2006). Furthermore, some of these satellites show traces of physical activity and renovation, possibly due to recent  geological processes. The geologic activity could include 
 endogenous activity such as viscous relaxation, volcanism, and/or tectonic or even atmospheric processes.  
Cratering itself  is a potential process in which the formation of large craters could remove small craters by  ejecta emplacement and seismic shaking. Also, exogenous processes, such as in fall from the E-ring or debris rings, could erase surface features.

Those physical processes would renew the satellite surfaces, erasing old or young surface features such as craters. The most striking case is Enceladus whose surface differs markedly from region to region and has present active geysers emanating from four parallel fractures in the south polar region, called ``tiger stripes'' (Porco et al., 2006). This region is a very active one where the surface is  modified, erasing  the craters (totally or partially). On Dione, in turn, Cassini observations reveal two different surfaces: the cratered plains (cp), heavily cratered, and the smoother plains (sp), with a lower cratering suggesting a younger surface. On Rhea, measurements by Cassini spacecraft detected a tenuous atmosphere of oxygen and carbon dioxide (Teolis et al., 2010).
Iapetus is the opposite case of Enceladus as it has heavily cratered plains  with large degraded basins, which indicates that its surface is ancient. What is more, the count of  small craters showed a  size distribution  indicative of crater saturation (Denk et al., 2010). 

Kirchoff and Schenk (2009) and (2010), hereafter KS09 and KS10, analyzed high-resolution Cassini images and obtained the number and size-frequency distribution (SFD) of craters for the mid-sized icy satellites. With their observed number of craters and the previous cratering rate estimations by Zahnle  et al. (2003), they calculated the surface ages of each satellite  for some crater diameters. 

There are a number of factors that must be taken into account in the analysis of the age of a satellite area. First, all possible impactor populations should be considered. The analysis of the images obtained by Voyager (Smith et al., 1981, 1982) implied that the satellites were struck by two different impactor populations: Population I, which produced a greater number of large craters (bigger than 20 km), and Population II, which produced a greater number of smaller craters (smaller than 20 km) (Smith et al., 1982). The origin or even the existence of both populations is disputed. Smith  et al. (1981, 1982) suggested that Population I was  the tail-off of a postaccretional heavy bombardment, while Population II has the form expected for collisional debris from the satellites or other orbiting debris.
 Horedt and Neukum (1984) concluded that cratering on Saturnian satellites is produced by heliocentric objects as well as by  planetocentric impactors.  On the other hand, Hartman (1984) noted that crater densities on heavily cratered surfaces throughout the Solar System are all similar due to a ``saturation equilibrium''. He argued that only one single population of heliocentric planetesimals  and their fragments had been recorded. 
Dobrovolskis and Lissauer (2004) studied the fate of ejecta from the irregularly shaped satellite, Hyperion, suggesting that it does contribute to Population II craters on the inner satellites of Saturn. However, those particles would produce craters with a different morphology than those produced by a heliocentric source. 

Moreover, consideration of the primordial situation is important for the study of the origin of craters. It is believed that the mass of the primordial trans-Neptunian zone, the source of Centaurs,  was $\sim 100$ times higher than the present one (Morbidelli et al., 2008). This mass might have been depleted by a strong dynamical excitation of the trans-Neptunian region. There were several models that described the mass depletion; for example,  in the ``Nice Model'' and its subsequent versions, the interaction between the migrating planets and planetesimals destabilized the planetesimal disk and scattered the planetesimals all over the Solar System before the time of the LHB (Tsiganis et al., 2005;  Levison et al., 2008). Accordingly, there was much primordial mass that struck the planets and their satellites very early in the Solar System history. It would be expected that this event  marked the surfaces of the satellites, mainly producing a great number of larger larger craters than at present. Those craters are probably the larger ones that can be observed on the satellites. Moreover, there are papers (Nervorny et al., 2003, 2007) that argue in favor of more primordial irregular satellites and a more active primordial collisional activity around the major planets. There would even be  populations of large irregular satellites and debris of catastrophically disrupted satellites that may have played  important roles in the history of the Saturn system (Dones et al., 2007).    

A primordial crater contribution to the  Saturn system  could also be connected with the crater saturation observed on some of the Saturnian satellites. The formation of craters, especially the larger ones, is a process that is effective in erasing small craters, not only through crater formation, but also by the ejecta blanket and seismic shaking. When a surface is so heavily bombarded that the formation of craters is equaled by the obliteration of craters, it reaches a crater saturation equilibrium. In this case, the density of craters on a surface does not change proportionally. This process is then critical to the determination of the  source of impactors, geological processes, ages, etc (Hartmann and Gaskell, 1997). Crater saturation equilibrium finally causes a distinctive cumulative size frequency distribution (SFD) of craters. Gault (1970) obtained a  $-2$ power-law, from a model based on first principles and laboratory experiments. Hartmann (1984) fitted a $-1.83$ power-law for the cumulative SFD to crater count data of the surface of various Solar System objects. Richardson (2009) developed a detailed cratered terrain evolution model and observed the way in which crater densities attain equilibrium conditions. He found that if the impactor population has a cumulative power-law slope of $< -2$,  crater densities reach a cumulative power-law of about $-2$; and if the impactor population has a cumulative power-law slope of $> -2$, crater density equilibrium values follow the shape of the production population. 
 Squyres et al. (1997) argued that cratering on a surface is a random process and as crater obliteration becomes more important, the spatial distribution of craters tends to be uniform. They used both the spatial distribution and size-frequency distribution of craters to study, with statistical techniques, how a cratered surface approaches saturation.  They found that at least $25 \%$ of the craters on Rhea and Callisto were destroyed by subsequent obliteration. Therefore, crater saturation is especially important on highly cratered surfaces that are in general old. 

In the supposedly young surfaces, as in the case of  Enceladus for example,  geological processes are erasing craters. However, such  processes compete with possible exogenous processes, like particle deposition on the surface that can fill in the craters, finally removing them. Mid-sized Saturnian satellites with the exception of Iapetus are embedded in the Saturn E-ring; therefore their particles cross the orbits of the satellites with some probability of impact. Enceladus's plumes are  thought to be the source and maintenance of  E-ring material (Hamilton and Burns, 1994, Kempf et al., 2010). A fraction of the plume particles escape to populate the E-ring but other fraction 
returns to Enceladus hitting its surface  (Kempf et al., 2010).  
Ingersoll and Ewald, (2011) estimated $(12 \pm  5.5) 10^8$ kg for the mass of particles in the E-ring and a particle lifetime in the E-ring of about 8 years. 
This material could be an exogenous source causing erasure of craters in the mid-sized Saturnian satellites.  

In a recent paper, Di Sisto and Zanardi (2013), hereafter DZ13, calculated the production of craters on the mid-sized Saturnian satellites produced by current Centaur objects coming from the Scattered Disk (SD) and plutinos, in the trans-Neptunian region, and compared these calculations with the Cassini observations made by KS09 and KS10. Also obtained was the current cratering rate on each satellite. In that paper, the authors concluded that since the number of observed small craters is lower than their calculated theoretical number, and because there are physical processes on at least some satellite surfaces that erode them, such difference is likely to be caused by those processes. It seems that the satellite surfaces were ``reset'' by those physical processes and were, therefore, younger and defined by the current geological processes. However, as mentioned, a number of factors has to be considered to test what processes are really taking place on the satellites and how these affect the age of their surfaces. 
Therefore, in this paper we analyze in detail our theoretical predictions about the current cratering  production to obtain satellite surface ages.    
 These age determinations are not only a first approximation but also  a way  to help determine the time scale of the dynamic and geological processes that are taking place on the mid-sized satellites of Saturn.

Since each satellite is a unique world, it is also interesting to test peculiarities and different processes on each satellite in our study.

\section{Age calculation}

For the calculation of the number of impact craters and ages on the  mid-sized Saturnian satellites all potential populations of impactors should be considered. Those impactors could be main-belt asteroids, Jupiter and Neptune Trojans, Jupiter Family Comets (JFCs) and Centaurs from the SD,  irregular satellites, planetocentric bodies and Nearly Isotropic Comets (NICs) (Dones et al. 2009). There is evidence, such as recent collisions and passages of comets by Jupiter and the observations of Saturn-crossing comets and Centaurs (Dones et al., 2009) as well as also previous work by Zahnle et al. 1998 and 2003, that Centaurs and JFCs dominate impact cratering in the Outer Solar System. The only truly unknown potential impactor population is the planetocentric one. This hypothetical source has been discussed by previous papers (see Sect. 1) but it has not been observed so far; its possible contribution or existence will be analyzed for each satellite with our results.    
  
Di Sisto and Zanardi (2013) calculated the production of craters on the mid-sized Saturnian satellites by Centaur objects from SD and plutinos. They found that the contribution of plutinos is negligible with respect to Scattered Disk Objects (SDOs). They used a method previously developed by Di Sisto and Brunini (2011), hereafter DB11. 
Both papers are based on the numerical simulation carried out by Di Sisto and Brunini (2007), who studied the evolution of SDOs for 4.5 Gyrs when they enter the Centaur zone, considering the current configuration of the Solar System. Di Sisto and Brunini (2007) studied  only Centaur objects when they enter the giant planet zone from the SD; they stopped their simulation when Centaurs enter the JFCs zone. Thus, the contribution of JFCs to the satellite cratering is not considered here. However, only a $\sim 22 \%$ of SDOs enter the JFC zone (Di Sisto and Brunini, 2007).  These objects were studied by Di Sisto et al. (2009) through a numerical simulation that included a complete dynamical-physical model. Only a fraction of the initial comets return to the Centaur zone, depending on the size of the comet (a JFC reduces its radius due to sublimation and splitting). We can then assume that this contribution should  be approximately one order of magnitude smaller than that of the Centaurs when they enter the Saturn zone from the SD.   

DZ13 and  DB11 used the output files of the encounters of SDOs  with Saturn when they enter the Centaur zone (from Di Sisto and Brunini, 2007)  to calculate the cumulative number of craters ($N_{c} (>D)$)  produced by current Centaurs on the  Saturnian satellites. It is generally accepted that the mass of the primordial trans-Neptunian zone, the source of centaurs,  was $\sim 100$ times higher than the present one (Morbidelli et al. 2008 ) and decreased to its present value in at most 1 Gyr (``Nice Model''). This early phase of the Solar System was not modeled in the numerical simulation by Di Sisto and Brunini (2007), which instead started after the initial mass depletion, when the Solar System began to stabilize, and essentially becomes like  the current configuration of the Solar System. As mentioned, this simulation was run for 4.5 Gyrs, since the exact duration of this early phase is not known; but it should be representative of the post initial mass depletion time, namely $\lesssim 4$ Gyr. Then, $N_c(>D)$ corresponds to the number of craters produced since the change in configuration of the Solar System predicted by the Nice Model ($\sim  4$ Ga).

In DZ13, we calculated the $N_{c} (>D)$ for each satellite and for two cases of the SFD of SDOs. That is, we considered a SFD of SDOs as a power law with a differential index $s_1 = 4.7$ for diameters $ d > 60$ km (from Elliot et al., 2005). The SFD breaks for  $ d \lesssim 60$ km, but given the uncertainty in the SFD for small objects, we considered two values of the differential index:  $s_2 = 2.5$ and $s_2 = 3.5$ (the Dohnanyi exponent). 
In DZ13, we also compared our theoretical number of craters with the number of observed craters obtained by  KS09 in  Figs. 2-7 of their paper. We noticed that our calculated number of craters for  $s_2 = 3.5$ is in general closer to the observed number by KS09. 
 However, for small craters, the observed number of craters is lower than the calculated one. Since in general there would be  physical processes on mid-sized Saturnian satellite surfaces that erode them, these processes  could erase small craters. The extent of this erasure is specific to each satellite because it depends on the process that is taking place on each one. It is possible, then, to account for this process by calculating the age of each satellite surface. 
The calculation of the age of each terrain is as follows: 
first, let us assume (on the basis of previous comments)that the Centaurs from the SD are the main source of craters in the current configuration of the Solar System and that, given the similarity with observations mentioned before, the calculated number of craters is the one obtained from the differential index:  $s_2 = 3.5$  of the SFD of small SDOs.
  This assumption can be considered  a first step in the determination of ages. Since Centaurs come mainly from the SD (Di Sisto and Brunini, 2007), they should be the most important current heliocentric source of craters. It should be highlighted that a possible planetocentric source of craters has been excluded since there are no estimations or calculations from this hypothetical source. 

From DZ13, we know the total number of craters on each satellite greater than a given diameter D, produced by Centaurs from the SD in the current configuration of the Solar System ($N_c(>D)$). We took the observed number of craters $N_{o}(>D)$ from KS09. Then, if $N_o(>D) > N_c(>D)$, we can state that the terrain has  preserved craters that were produced at earlier times of the Solar System, or that there have been other important sources of craters (such as a planetocentric source), and/or there is no current geological activity  that could erode the craters. 
But if $N_o(>D) < N_c(>D)$,  there are $N_c(>D)-N_o(>D)$ craters that have been erased and then there must  be a geological process taking place on the satellite surface. The comparison between the calculated and the observed numbers can be seen in Figs. 2-7 in DZ13, where it is noticeable that there is a range of crater diameters (different for each satellite) for which  $N_o(>D) < N_c(>D)$. Therefore, those terrains could have a  geological activity that has eroded the craters, and the surface would be young.   

\begin{figure}[t!]
\centerline{\includegraphics*[width=0.9\textwidth]{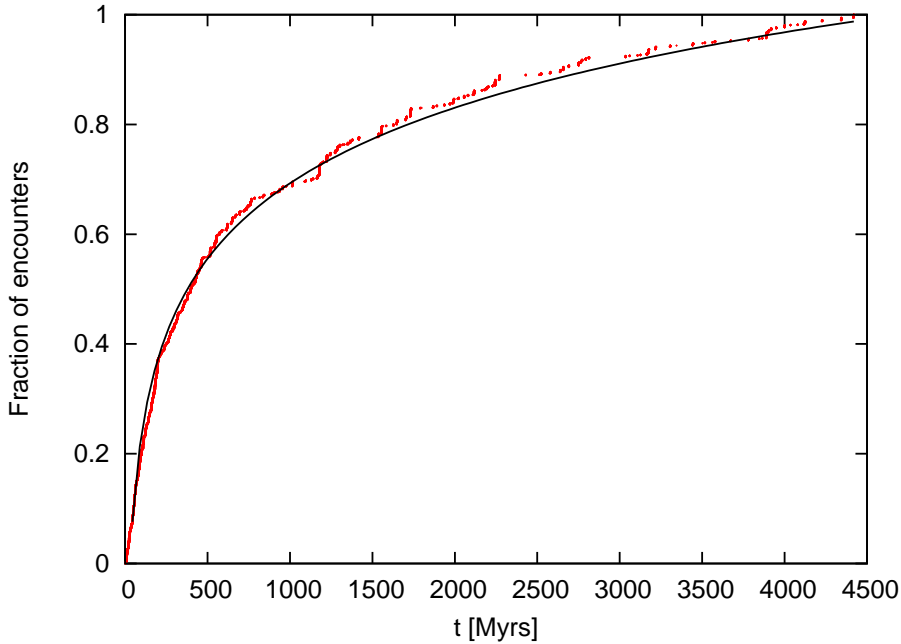}}
\caption{Fraction of encounters of SDOs with Saturn  versus time and the fit to the data.}
\label{enc}
\end{figure}

To calculate the age of those surfaces, the  temporal dependence of cratering was used. 
According to the method  developed by DB11, the cumulative number of craters produced by Centaurs is proportional to the number of encounters of Centaurs with Saturn. Then, the temporal dependence of cratering is equal to that of encounters with Saturn (see Eqs. 2-7 in DB11). In Fig. \ref{enc} of this  paper, the fraction of encounters with Saturn (number of encounters for a given time with respect to the total number of encounters) was plotted as a function of time and found that the whole plot could be fitted by a log-function given by: 

\begin{equation}
F(t) = a \,\, log \, t + b, 
\label{f} 
\end{equation}

where $a=0.198406 \pm 0.0002257$ and $b=-3.41872 \pm 0.004477$.

Then, the cumulative number of craters on a given satellite depending on time can be obtained from:

\begin{equation}
N_c(>D,t) = F(t) N_c(>D), 
\label{nct} 
\end{equation}

Differentiating this equation with respect to time, we obtain the cratering rate over time from: 

\begin{equation}
\dot C(>D,t) = \frac{a}{t}  N_c(>D), 
\label{ct} 
\end{equation}
 
Zanhle et al. (1998) calculated cratering rates on the Galilean satellites considering the contribution of several heliocentric sources.  They also calculated the age of the satellite surfaces. To such end, they considered that the cratering rate on each satellite decreases with time as $t^{-1}$ from a numerical simulation developed by  Holman and Wisdom (1993) who found that a collisionless Kuiper Belt dissipates as $t^{-1}$. This is the same behavior of our calculated cratering rate of Eq. \ref{ct}, since the original simulation by Di Sisto and Brunini (2007) deals  with the dynamical evolution of TNOs. Then, we follow a development analogous to Zanhle et al. (1998) to find the age of a satellite surface but now considering Eq. (\ref{ct}) for the cratering rate. Thus, the estimated cratering timescale or the  ``age'' of the surface  $\tau$ would be given by:

\begin{equation}
\tau(>D) = t_0 (1-e^{-\frac{N_o(>D)}{a N_c(>D)}} ) , 
\label{age} 
\end{equation}
where $t_0 = 4.5$ Gyrs is the age of the Solar System.

$\tau$ represents the model age of the surface for a given diameter, because it is a measure of  the cratering timescale, taking into account the production of craters and the erosion of the terrain.   

In the following sections, we  show our results for the mid-sized Saturnian satellites.

\section{Results} 
\label{r}

\begin{figure}[t!]
\centerline{\includegraphics*[width=0.9\textwidth]{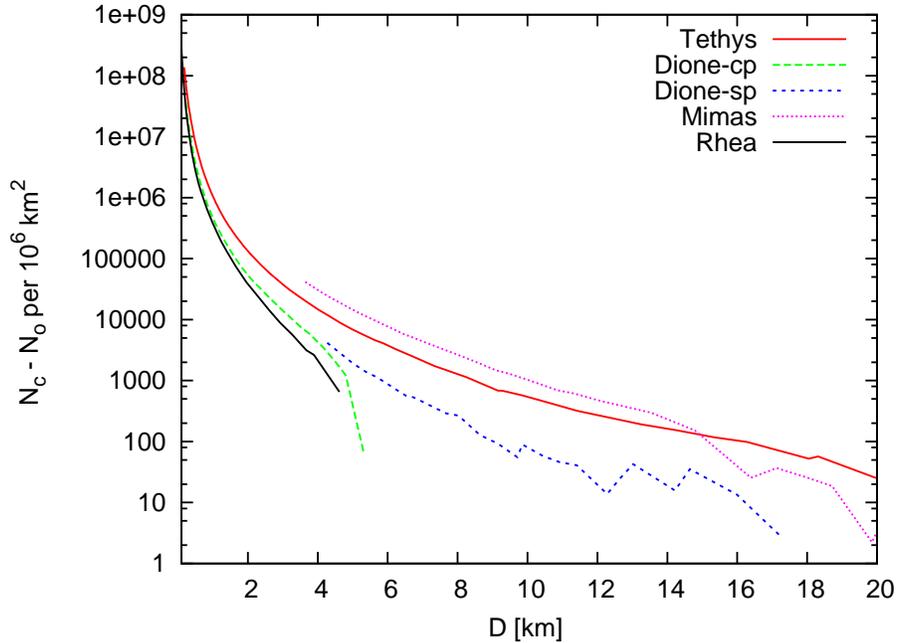}}
\caption{Difference between the cumulative number of theoretical craters and observed craters for Tethys, Dione, Mimas and Rhea according to D.}
\label{ncno}
\end{figure}

As mentioned in the previous section, considering that Centaurs are the current main source of craters, if $N_o(>D) < N_c(>D)$, there are $N_c(>D)-N_o(>D)$ craters that could have been erased. This difference can be plotted against $D$ to get an  understanding of the magnitude of the process on each satellite. In Fig. \ref{ncno}, we plot $N_c(>D)-N_o(>D)$ versus $D$ for Mimas, Tethys, Dione and Rhea considering $N_o(>D)$ from KS09. For all those satellites, the smaller $D$ is, the greater the difference between observation and our model. Notice also the different slopes of the curves. Tethys and Mimas have more flattened curves meanwhile the curves for Dione and Rhea are steeper. Therefore, for Tethys and Mimas, there seems to be a process that is acting for a wider range of diameters than for Rhea and Dione.

\begin{figure}[t!]
\centerline{\includegraphics*[width=0.9\textwidth]{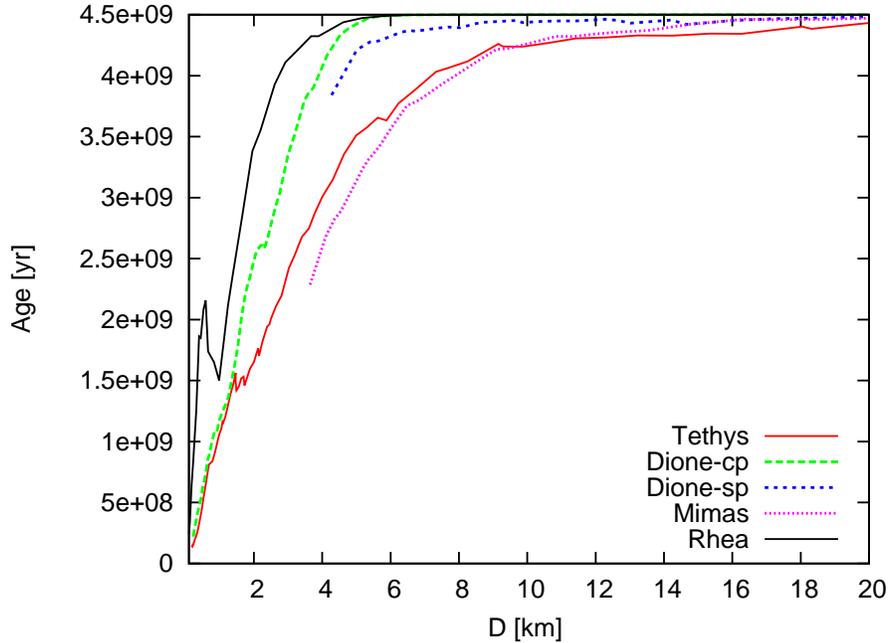}}
\caption{Age of the surface of Tethys, Dione, Mimas and Rhea according to D.}
\label{todos}
\end{figure}

We calculate the age depending on D from Eq. \ref{age} and plot the results for Mimas, Tethys, Dione and Rhea in Fig. \ref{todos}. These age-curves  correspond to the terrains analyzed by KS09.     
In all the curves, there seems to be a noticeable correlation between older age and diameter; that is, smaller craters are younger than greater ones. 
As can be seen, only small craters would be young or, at least, younger than 4.5 Gyrs old. 
This means that there could be a geological or other physical process that gradually erases craters and modifies the satellite surface; naturally, smaller craters are cleared before greater ones. Furthermore, the slopes of the age curves are different for each satellite. This fact could imply either that each satellite presents different geological processes or exogenous erasing agents (like deposition of E-ring particles), or at least, that each process has a different temporal scale. It could also be possible that on a given satellite surface the equilibrium crater saturation  had been reached. 

There also exists a correlation between the size range or erased craters and  distance to Saturn. In the ranges of the diameters plotted, Mimas, the innermost mid-sized satellite located in the inner edge of the E-ring, has a wider range of diameters of erased craters than Tethys, which is farther away from Saturn. Then follows Dione in distance to Saturn,  and finally Rhea, at the end of the E-ring. This correlation is very interesting, and potentially implies that one can associate a modification surface agent that depends on the distance to the planet.

Using Eq. \ref{age}, there is a certain diameter $D$ from which $N_o(>D) $  gets greater than  $N_c(>D)$ and then $ e^{-\frac{N_o(>D)}{a N_c(>D)}}$  approaches 0. Therefore, the calculated ages trend asymptotically to the age of the Solar System and  for those diameters, the surface would be old. As mentioned in the previous section, when  $N_o(>D)  > N_c(>D)$,   there are more craters on the surface of the satellite than our calculated number. Since  $N_c(>D)$ represents the crater contribution from Centaurs in the current configuration of the Solar System, there are craters that were produced earlier in the Solar System history and/or there is another source of craters, such as  planetocentric objects. The limit diameters beyond which $N_o(>D) \gtrsim N_c(>D)$ (or before which $N_o(>D) < N_c(>D)$) are presented in table \ref{dlim}; it can be seen that those limit diameters are different for each satellite and might be associated with different kinds of processes. Therefore, we cannot evaluate other crater sources for all satellites in general, but we can analyze all our results for each satellite in separate subsections.   

\begin{table}
\begin{minipage}[t]{\columnwidth}
\caption{Limit diameter beyond which $N_o(>D) \gtrsim N_c(>D)$)  }
\label{dlim}
\centering
\renewcommand{\footnoterule}{}  % to avoi
\begin{tabular}{lc} \\
\hline
Satellite  &  $D$ [km]   \\
\hline
Mimas    &   15        \\
Enceladus &  none      \\
Tethys   &  25   \\
Dione-cp  &  5     \\
Dione-sp  &  17    \\
Rhea  &  5     \\
Iapetus &  2   \\
\hline

\end{tabular}
\end{minipage}
\end{table}

\subsection{Iapetus}

Iapetus is Saturn's third-largest moon after Titan and Rhea. Its orbit around Saturn has a semimajor axis of $3.56 \times 10^6$ km ($\sim  59$ Saturn radii), low eccentricity and a slightly high inclination of $8.3^{\circ}$. 
Cassini images revealed a near equatorial ridge system that extends for more than $110^{\circ} $ in longitude and that rises more than 20 km above the surrounding plains (Porco et al., 2005). The ridge morphology is consistent with 
 an endogenous origin (Ip,  2006). Giese et al. (2008) found substantial topography on Iapetus' leading side from Cassini images  with heights in the range of $-10$ km to $+13$ km.
A distinctive feature of Iapetus  is its global albedo dichotomy. The leading side is dark with  albedo values of $\sim 0.04$ on a roughly elliptical area (named Cassini Regio after  the astronomer Jean Cassini who first noticed it); the trailing side, in turn, is relatively bright with albedos of $\sim 0.6$. Two classes of theories have been suggested to account for the origin of this dichotomy:  the endogenous ones (Smith et al., 1981, 1982) and the exogenous theories which suggest that debris or dust material from the external moons of Saturn impact the dark hemisphere (Cruikshank et al., 1983; Buratti and Mosher, 1995). A model including both endogenous and exogenous causes was developed by Spence and Denk (2010) which demonstrates that the dichotomy can be explained by runaway global thermal migration of water ice, triggered by the deposition of dark material on the leading hemisphere. From a dynamical point of view, the routes of possible dust impactors on Iapetus were studied by Leiva and Briozzo (2013). They analyzed  low-energy incoming dust particles and obtained their distribution on the surface of Iapetus.   
There is also a global color dichotomy found on Iapetus (Denk et al., 2010), the leading side being  redder than the trailing side. Spencer and Denk (2010) provided evidence for an exogenous origin for the redder leading-side parts and suggested that it initiated the thermal formation of the global albedo dichotomy.  

\begin{figure}[t!]
\centerline{\includegraphics*[width=0.9\textwidth]{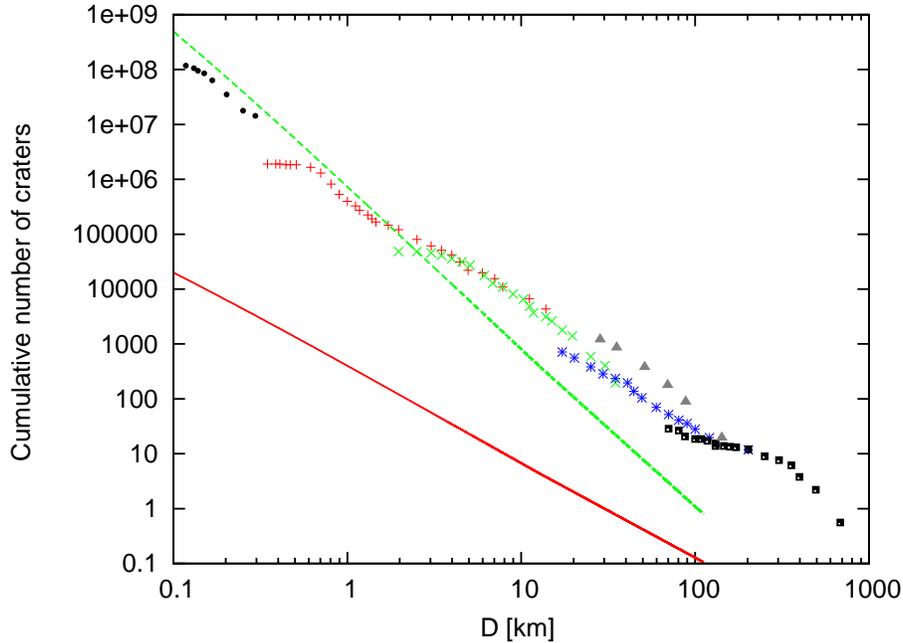}}
\caption{Cumulative number of craters of Iapetus with respect to D. Points represent the number
of craters obtained by Denk et al. (2010) and Plescia and Boyce (1983) (the gray ones), the red curve represents the cumulative number of craters obtained by our model for $s_2 = 2.5$, and the green one for  $s_2 = 3.5$.}
\label{iapetus}
\end{figure}

Iapetus is heavily cratered, with degraded basins with diameters nearly the size of Iapetus's radius, which indicate that both the bright and the dark areas are ancient. Plescia and Boyce (1983) analyzed Voyager 2 images, determining the number of craters at large diameters (because of low image resolution) that indicated that the bright terrain is an ancient surface that dates to the LHB (Plescia and Boyce, 1985).  Denk et al. (2010) plotted the cumulative crater size-frequency distribution  of Iapetus  $N_o(>D) $, combining five individual measurements (from Cassini images) that we reproduce in Fig. \ref{iapetus} (extrapolating their counts to the complete Iapetus surface) together with Plescia and Boyce (1983) counts. In this figure, we also plotted the cumulative number of craters on Iapetus greater than a given diameter D, produced by current Centaurs from the SD ($N_c(>D)$) calculated from our model. 
As mentioned, we assumed a SFD of SDOs that  breaks for $ d \lesssim 60$ km  considering two values of the differential index:  $s_2 = 2.5$ and $s_2 = 3.5$. The two resulting curves are plotted in   Fig. \ref{iapetus}. As  noticed by  DZ13 for the other satellites,  our theoretical number of craters for  $s_2 = 3.5$ (green curve) is  closer to the observed number. Moreover, both  $N_c(>D)$ and  $N_o(>D) $ are similar for  $D \lesssim 2$ km and for  $D \gtrsim 2$ km $N_o(>D) > N_c(>D)$. We already noticed in DZ13, comparing our results with those of KS10, that  $N_o(>5$ km) $ > N_c(>5$ km), which is  consistent with an old surface. 
 Denk et al. (2010) found that for $D \lesssim 5-10$ km, the crater cumulative SFD  follows a $-2$ power law that corresponds to crater equilibrium saturation. In fact, by fitting power-laws of the form  $N_o(>D) = c D^b$ to the observed crater distributions plotted in Fig. \ref{iapetus}, we found for each terrain and ranges of diameters,  different values of b. These are shown  in Table \ref{denk}.  

\begin{table}
\begin{minipage}[t]{\columnwidth}
\caption{ Power-law indexes of the observed cumulative crater distribution obtained by fitting to the observed craters by Denk et al. (2010), Plescia and Boyce (1983) and  KS10}
\label{denk}
\centering
\renewcommand{\footnoterule}{}  % to avoi
\begin{tabular}{lrccc} \\
\hline
Region      &     $D$-range [km]  & b  \\
\hline
\hline
Anti-Saturn hemisphere &  $5<D<35$     & -2.2     \\
                 & $8<D<35$    & -2.36    \\
                 &  $10<D<35$    & -2.47     \\
\hline
Center of basin Engelier &  $2<D<14$    &  -1.66      \\
\hline
Large craters on trailing side &  $17<D<202$ & -1.7      \\
\hline
Basins (global) &  $70<D<350$   &   -0.92    \\
                & $350<D<684$   & -3.48  \\
\hline
Iapetus-dark (KS10)  &  $4<D<80$   & -2.66      \\
\hline
Iapetus-bright (KS10) &  $4<D<65$   & -1.7    \\
\hline
Iapetus-bright (PB83) &  $20<D<142$   & -2.02    \\
\hline
\end{tabular}
\end{minipage}
\end{table}

In the anti-Saturn Hemisphere region, Iapetus-dark and even large basins, there seems to be a tendency of a steeper slope for greater sizes. Therefore, considering that for $D \lesssim 2$ km, the crater distribution is saturated, a  decrease of b for larger craters  might suggest that larger craters are not likely  saturated. If this interpretation were correct, then the crater distribution would reflect the impactor distribution only for large craters. 
Our theoretical crater distribution $N_c (>D)$ with $s_2 = 3.5$ (an impactor population with a cumulative SFD  with an index of $2.5$) has a power-law index of $b = -2.976$ for $D \gtrsim 0.08$ km. Hence, only for dark terrain and $D>10$ km of the anti-Saturn hemisphere, the observed crater distribution approaches our theoretical crater distribution. 
On the bright hemisphere, however, the observed crater power-law index is $-1.7$ from Cassini counts and $-2.02$ from Voyager counts, a much flatter slope than the theoretical law.

On the other hand, from Fig. \ref{iapetus} we note that for $D \gtrsim 5$ km, $N_o(>D) > N_c(>D)$ with a difference $\sim 10$ to $\sim 30$ times greater.  Besides, according to our calculations, the greatest crater produced by current centaurs has $D \sim 100$ km, but there is observed a significant number of craters and basins greater than this diameter. Therefore,  since our model represents the present contribution to cratering, and a current planetocentric population of big objects near Iapetus is not expected, the main crater source of Iapetus must be primordial, which is  consistent with previous studies (Plescia and Boyce, 1983, 1985). As mentioned, it is believed that the mass of the primordial trans-Neptunian zone was $\sim 100$ times higher than the present one (Morbidelli et al., 2008), in  agreement with the differences found between observation and present modeled contribution. 

Therefore, from the comparisons of our model and observations, we can conclude that big craters on Iapetus are old and connected with the primordial epoch of the Solar System. As the crater size decreases,  it seems to approach saturation until $D \lesssim 2$ km-craters, where saturation is complete (Denk et al., 2010). This crater saturation does not allow us to evaluate the possibility of another current source of craters such as planetocentric small objects.  The different observed power-laws between bright and dark terrains, which may be an artifact of analyzing small areas (Kirchoff and Schenk, 2010), cannot be explained by our model. If real, this difference  could be connected to primordial times or also to probable apex-antapex asymmetry (Zhanle et al., 2003). 

Furthermore, an interesting  result obtained from our model is the recent production of small craters. Denk et al. (2010) identified brighter-than-average craters on highest-resolution images on the dark side, with diameters of up to $\sim 200$ m, the brightest one being  the crater Escremiz, which is $\sim 4$ times brighter than its surroundings. They estimated crater ages assuming that brightness and age are correlated  and that the darkening process acts uniformly for all fresh craters. They also stated that a new crater is $\sim 10$ times brighter than the surrounding dark terrain and that it declines to half brightness in $\sim 10000$ years. Therefore, crater Escremiz is estimated to be  $\sim 10000$ years old. 
Moreover, they estimated that on any of Iapetus' dark hemisphere, the largest crater with an age similar to Escremiz should have $D \sim 200$ m, and that slightly more than 100 craters of 60 m or larger and with an age similar to or younger than Escremiz should exist.   
From our model, it is possible to calculate the time the production of craters with a diameter greater than a given $D$. From  Eqs. \ref{nct} and \ref{f}  and considering that Cassini Regio covers $\sim 40 \%$ of Iapetus surface, we found that the last crater greater than $ D$ was produced $\Delta t$  years ago  given by:

\begin{equation}
\Delta t(>D) = t_0 (1-e^{-\frac{1}{a 0.4 N_c(>D)}} ) , 
\label{age} 
\end{equation}

Then, the last crater with $D > 60$ m in Cassini Regio was produced $\sim 30$ years ago, and  the last 100 craters with $D > 60$ m in Cassini Regio were produced in the last $\sim 3000$ years. The last  crater with $D > 200$ m in Cassini Regio was produced $\sim 800$ years ago. Those ages are somewhat smaller than the Denk et al.'s (2010) predictions. This  might suggest that our calculations of the contribution of Centaurs to the recent cratering are more numerous than expected. However, these differences could be related to the probable apex-antapex asymmetry (Zhanle et al., 2003), which is not considered in this model, or to the intrinsic errors of the current number of SDOs at small sizes or even to errors in  Denk et al.'s (2010) model of age-brightness correlation.

\subsection{Mimas}

Mimas is the innermost  mid-sized Saturnian satellite and its orbit is located within the inner edge of the E-ring. Mimas has a diameter of $394.4$ km, and has a density of $1.149 gr/cm^3$ slightly higher than water. 
 Mimas' surface is fully covered with impact craters and is one of the most heavily cratered bodies in the Solar System. This satellite has a large impact crater of 139 km in diameter (one third the diameter of Mimas) called Herschel. Schmedemann and Neukum (2011) also suggest evidence for two highly degraded large craters, one of which is 153 km in diameter. After calculating the production of craters by Centaur objects on Mimas, DZ13 found that the largest crater produced by Centaurs has a diameter of 113 km, whose size is similar to Herschel crater.
Lastly, KS10 used high resolution images from Cassini to record surface impact craters of the surface and they found a high density of craters with diameters $\sim$ 10 $<$ $D$ $<$ 30 km. 
Plescia and Boyce (1982) obtained crater counts in different areas of Mimas from Voyager images. They distinguished an equatorial area near Hershel (of low resolution), and south polar and adjacent areas which generally lack craters  $>20$ km.   

\begin{figure}[t!]
\centerline{\includegraphics*[width=0.9\textwidth]{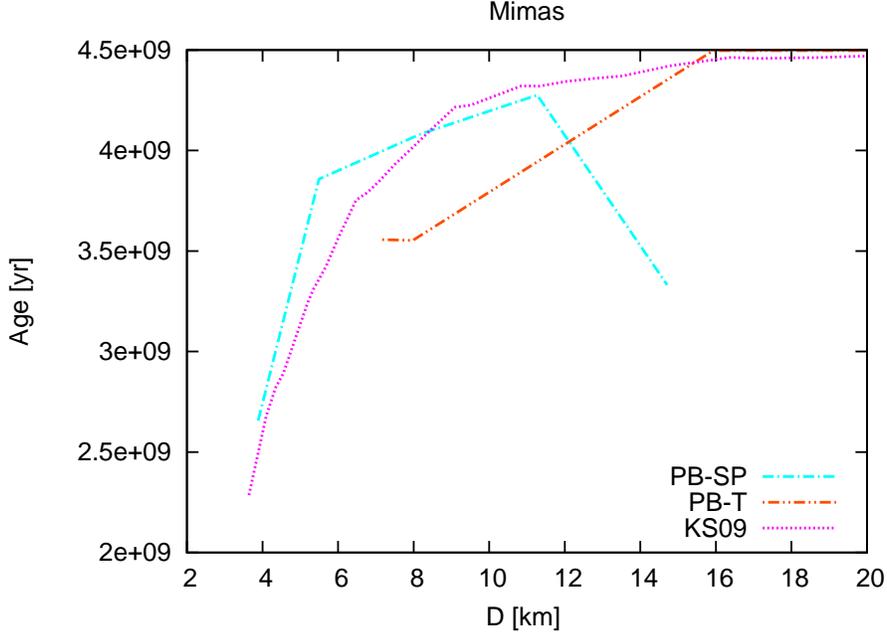}}
\caption{ Age of the south polar (PB-SP) and intermediate latitudes regions (PB-T) from Voyager and 
  Cassini images (KS09) of Mimas with respect to D.}
\label{mimas}
\end{figure}

Buratti et al. (2011) have failed to find plume activity on the satellite and  
 concluded that there could only be low-level geologic activity. They  also limited this plume activity to an upper level of one order of magnitude less than the production for Enceladus. 
Measurements of Mimas's forced librations by Tajeddine et al. (2014) confirms the  libration amplitudes calculated from the orbital dynamics, with the exception of one amplitude that depends on Mima's internal structure. They suggested that a non-hydrostatic core or a subsurface ocean are the two possible interior models of Mimas that are consistent with their observations.    
 It is evident that age can be calculated only if there exist measures of the number of observed craters. For example, in the case of Mimas, as there are no counts for craters with $D \lesssim 3.5 $ km, the age-curve in Figs. \ref{todos} and \ref{mimas} begins at this diameter. We calculate the age depending on D from Eq. \ref{age}, considering  crater counts of the  equatorial (E), south polar (SP) and intermediate latitudes regions (T) from  Plescia and Boyce (1982), and also crater counts by KS09 from Cassini images of half of Mima' surface. We plot the results in Fig. \ref{mimas}. Crater counts for the equatorial zone in Voyager images are given only for large diameters due to low resolution, and the age is  nearly $4.5$ Gyrs old.        

The south polar terrain of Voyager images has a similar age curve to Cassini images for $D\lesssim 12$ km, but for larger craters, the south polar terrain seems to be younger, probably due to the lack of craters  $>20$ km found by Plescia and Boyce (1982). Since on the one hand, Cassini counts include almost half of Mimas' surface and  have a better resolution and on the other hand, SP area is small, it is possible that this drop of the SP age-curve is very local.      
 For T-area and Cassini images, craters with $D \lesssim 15$ km are younger than 4.5 Gyrs old, and smaller craters might be even younger. For example, craters with $D \lesssim 6 $ km  are younger than $\sim 3.5$ Grys old. In Fig. \ref{ncno}, there are $\sim 8000$ craters per million km$^2$ with $D \gtrsim 6$ km that could have been erased from the surface in the age of the Solar System.  Therefore, one possibility is that, effectively, small craters could have been erased due to particle depositions from the E-ring or even from a low-level geologic activity that  should not be discarded on Mimas. 
However, since Mimas is heavily cratered, another possibility is that for $D \lesssim 20$ km, crater density has reached a saturation equilibrium. In fact, KS09 found that the power-law index of the observed cumulative crater SFD is  $-1.548$ for $ 4$ km$ < D < 10 $ km  and  $-2.12$,  for $ 10$ km $ < D < 20$ km, both indexes near to the saturation equilibrium condition.
The slope behavior of the age-curve is similar to Tethys and both curves are below the other two curves of  the  satellites plotted, which shows an erasure process or saturation that acts on a wider range of crater sizes on those satellites. 

Because of the recently suggested potential for activity on Mimas, we cannot discard an erasure process accounting  for the paucity of small craters suggested in our model. However, the more conservative hypothesis would be  saturation equilibrium for craters less than $20$ km in diameter.

\subsection{Tethys}

Tethys is  cold and airless, 1072 km in diameter, and  very similar in size and aspect to  Dione and Rhea. As it has a density of $0.985 gr/cm^3$ (Thomas, 2010), very similar to liquid water,  it is probably almost entirely  composed of water ice and a small amount of rock. In fact, spectral observations of Tethys' surface show no other component than $H_2O$ ice (Emery et al., 2005). 
This satellite  has two main features on its surface: one of them is a great impact crater known as Odysseus, whose diameter is $\sim$ 450 km, and the other one is a great graben called Ithaca Chasma, whose length is $\sim$ 2000 km. The interior of Odysseus seems to be younger than the rest of Tethys' surface. In general, the surface of Tethys seems to be old and highly cratered, showing little geologic diversity.

Royer and Hendrix (2014) present observations from the Ultraviolet Imaging Spectrograph subsystem of Cassini for Mimas, Tethys and Dione. They found that Tethys and Dione have a leading hemisphere brighter than their trailing hemisphere at far-UV wavelengths, and Tethys shows a narrower opposition effect, reflecting a more porous surface. They attribute this structure to  an intense bombardment by E-ring.   

\begin{figure}[t!]
\centerline{\includegraphics*[width=0.9\textwidth]{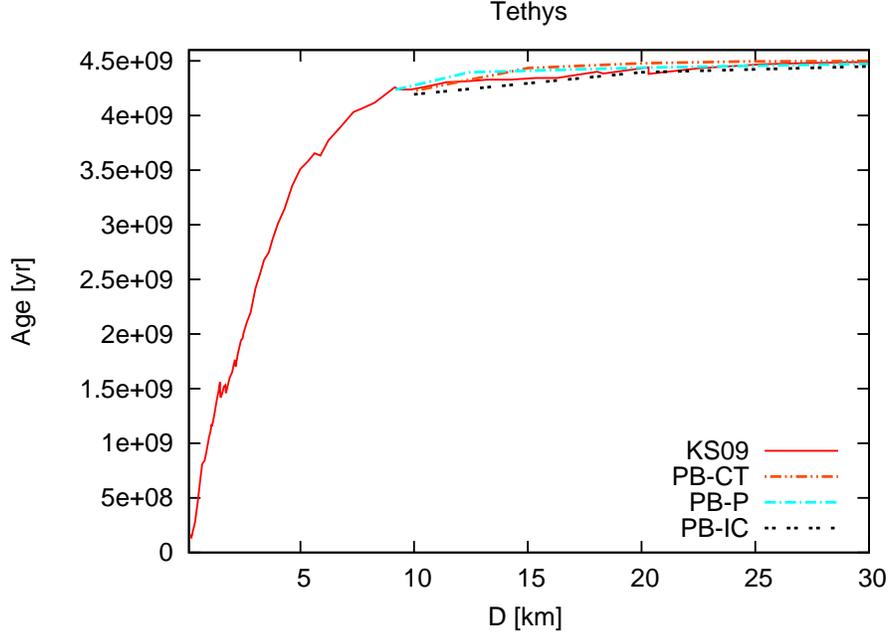}}
\caption{ Age of the Ithaca Chasma (PB-IC) interior terrain,  plain units (PB-CT) and (PB-P) from Voyager images and   Cassini images (KS09) of Tethys with respect to D.}
\label{tethys}
\end{figure}

There are crater counts on Tethys' areas from Voyager images (Plescia and Boyce, 1982 and 1983) and from Cassini images (KS09). From those observed crater counts and our model, we calculate the age depending on D from Eq. \ref{age} for those terrains. The two bands, along the west side of the terminator observed by Voyager and studied by Plescia and Boyce (1982) have ages  near $4.5$ Gyrs, consistent with the  high craterization observed. The other four zones analyzed by Plescia and Boyce (1983) have crater counts for $D > 10$ km. The heavily cratered terrain near $60^{\circ}$ longitude is $4.5$ Gyrs old. The remaining three areas that include two plain units (CT and P) and the interior of Ithaca Chasma graven (IC) have craters with $D \lesssim 15-20$ km that are younger than 4.5 Gyrs old, which is consistent with the age calculated by Plescia and Boyce (1985). Crater counts on Cassini images performed by KS09 were done for $\sim 0.2 < D \lesssim 90 $ km and they have similar ages to plain units and IC for $ D \gtrsim 10$ km. Ages depending on D for CT, P, IC  and KS09  terrains are plotted in Fig. \ref{tethys} and the  heavily cratered terrains have been omitted for clarity.

In Fig. \ref{todos}, we plot the age-curve of Tethys only for KS09 crater counts, since it has a wider range of diameters and a better resolution. It  shows an age-curve similar to that of Mimas for the range of diameters plotted. Craters with $D \lesssim 20 $ km could  be young and, like the other satellites,  smaller craters are younger than larger ones.  KS09 and KS10 calculated the cumulative slopes of the SFD of craters and found that for $0.2$ km $ < D < 10$ km  the cumulative slope is $-1.728$,  and for $ 10$ km $ < D < 60$ km is $-2.2$. Then, for small craters those distributions could be close to  the saturation equilibrium, especially for $D \lesssim 10$ km. Craters  greater than  $D \sim 20 $ km seem not to be affected by the erasing process or saturation,  and thus the surface has preserved them, as can also be seen in Fig. 5 of DZ13. Therefore, those craters should be primordial.  

Like on Mimas, observations by Buratti et al. (2011) do not show plume activity, and the upper limit of them are also one order of magnitude less than the production on Enceladus. However, this does not rule out some possible activity. Since there appear to be small craters that would have been erased, there could, in fact,  be  some kind of geological activity (even plume activity at a low level, which has not been detected yet) that is acting on the satellite, and/or also bombardment of the E-ring flux, as was suggested by Royer and Hendrix (2014). As can be seen in  Fig. \ref{ncno}, this process  should be able to erase, for example, $\sim 14000 $ craters per $10^6$ $km^2$ with $D\gtrsim 4$ km or $\sim 10^6  $ craters per $10^6$ $km^2$  with  $D\gtrsim 1$ km in 4.5 Gyrs.

However, as mentioned before, crater saturation in small crater sizes is the most conservative hypothesis for the paucity of observed counts.

\subsection{Dione}

Dione has a diameter of 1122.8 km and a mean density of $1.478 gr/cm^3$. 
Its surface is uneven, as it presents heavily cratered terrains with craters greater than 100 km in diameter, as well as  smooth plains (Smith et al., 1982; Plescia and Boyce, 1982). It shows a marked asymmetry between its leading and trailing hemispheres, with the leading side  brighter than the trailing one. Royera and  Hendrix (2014) analyzed the photometric properties of Dione revealing a more absorbent surface. They ascribe this to the fact that Dione is less intensely bombarded by E-ring grains and thus it should have less fresh bright water-ice on its surface;  and/or to that  an exogenous darkening agent could be acting simultaneously on  the trailing side.

In addition, Buratti et al. (2011) analyzed spectral observations of Dione in search for plume activity and found that the upper limit of water vapor production  is two orders of magnitude smaller than that of Enceladus, suggesting that this world is inert. However, plume activity cannot be absolutely ruled out because it can be under the detection limit. 
Further analysis by the Cassini magnetometer suggests that Dione has a tenuous atmosphere of transient origin (Simon et al., 2011). Future observations by Cassini could possibly confirm the existence of an atmosphere and/or some plume activity.

\begin{figure}[t!]
\centerline{\includegraphics*[width=0.9\textwidth]{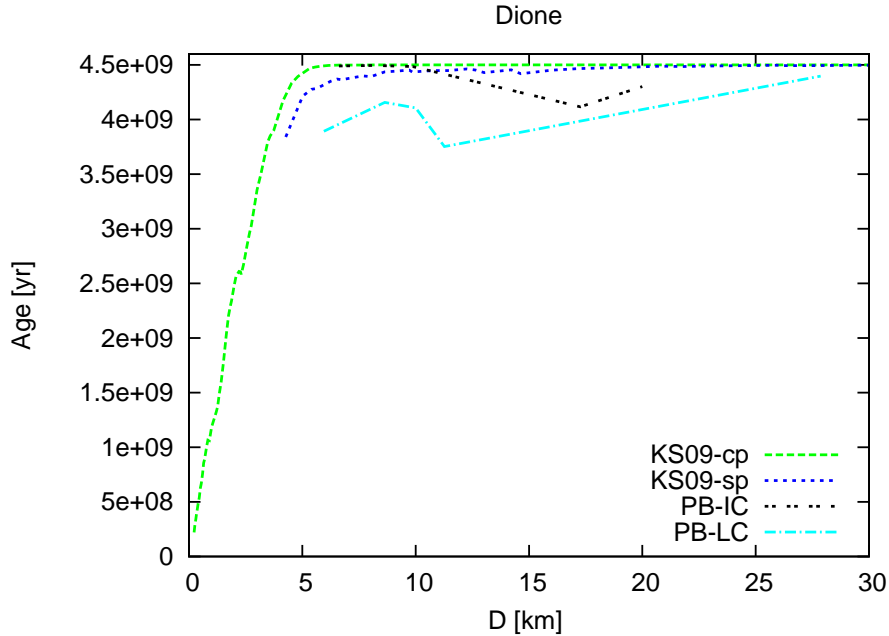}}
\caption{ Age of an intermediate crater plains unit (PB-IC) and a smooth plains unit (PB-LC) from Voyager images and  cratered plains (cp) and  smooth plains (sp) from  Cassini images (KS09) of Dione with respect to D.}
\label{dione}
\end{figure}

Plescia and Boyce (1982) analyzed  three terrain types observed by Voyager: a heavily cratered rough terrain (HC), an intermediate crater plain unit (IC) and a smooth plain unit (LC). They obtained cumulative size frequency distributions of craters for those terrains, which we used in our model to calculate surface ages. The HC area has crater counts for $\sim 10 < D \lesssim 100$ km and is $4.5$ Gyrs old. Ages  depending on D of IC and LC plain units are shown in Fig. \ref{dione}. The IC unit has crater counts for $6 \lesssim D \lesssim 20$ km and it seems that craters between 15 and 20 km are slightly younger than 4.5 Gyrs old. The LC terrain would be younger than 4.5 Gyrs old for all the diameters observed. This is consistent with the low crater density of LC and IC plains with respect to HC terrain found by Plescia and Boyce (1982) and also with the younger ages found by Plescia and Boyce (1985) for those IC and LC plains.

On Dione there are two different surfaces recognized by KS10 from Cassini observations. The cratered plains (cp), which are heavily cratered, presenting  records of primordial bombardment, and the smooth plains (sp) with less cratering, suggesting a younger surface. In Figs. \ref{dione} and  \ref{todos}, we plot the age with respect to  D for both terrains on Dione. 

In these figures we can see that the observed cp surface portion of Dione is generally old, but craters with $D \lesssim 5 $ km could  be young, and  smaller craters would be younger than larger ones. This could be the result of a process in which craters have been gradually erased.  This possible erasing process does not appear to affect craters greater than $\sim 5$ 
km. Although there is no data for  $D \lesssim 4 $ km for sp, the smooth terrain has a lower amount of craters than the cp surface, and the corresponding age curve is below the cp-age curve. For example, 
 for craters with  $D \gtrsim 4 $ km our model suggests that sp has an age of $3.8$ Gyrs old while the cp is 4 Gyrs old. Indeed, since the age of sp for  $D \gtrsim 8 $ km is near $4.5$ Gyrs old  (Fig. \ref{ncno}) and the number of calculated craters is greater than the observed one, this suggests a very slow erasing process for $D \gtrsim 5 $ km, which is not present in the case of cp. Potential erasing process/es on Dione would need to  be able to erase in sp terrain,  $\sim 5500 $ craters per $10^6 km^2$  of $D \gtrsim 4$,   while in cp terrain there would be $\sim 4500 $ craters per $10^6$ km$^2$ of $D \gtrsim 4 $ erased in 4.5 Gyrs. 
Comparing age estimations between Voyager and Cassini plains, it is evident that LC plains would be younger than smooth plains, at least in the range of diameter observed by both missions. It could be that LP plains were in fact young due to, for example, the internal activity which resurfaced portions of the surface of Dione, as was suggested by Plescia and Boyce (1985). However, Cassini images have a resolution  of 1 to 2 orders of magnitude higher than Voyager images, so the comparison between age calculation with the data of both missions  has to be taken with caution.   

On the other hand, like on the other satellites,  KS10 calculated the cumulative crater SFD  for cratered plains and obtained a cumulative slope of $b = -1.640$,  for $ 0.25 $km $ < D < 4$ km, $b = -1.166$ for $ 4$ km $ < D < 10$ km, $b = -2.31$ for $ 10$ km $< D < 30$ km,  and $-2.9$ for $ 30 $km $ < D < 150$ km.
Therefore, the crater SFD of cp is similar to our model for craters greater than $30$ km, but for smaller sizes the slope of the SFD is getting more flattened, similar to what occurs for Iapetus, indicating  saturation equilibrium for $D \lesssim 10 - 30$ km. For the smooth plains  a similar behavior of the crater SFD has been found.
 Plescia and Boyce (1982) found a cumulative slope of $b = -2$ for the heavily cratered terrain, indicating  saturation equilibrium, and a cumulative slope of $b = -4$ for the younger units, more similar to our model. 

Then, crater saturation could also be responsible for the paucity of small craters on Dione although there is morphological evidence that small craters in the sp have also been affected by endogenous resurfacing (Schenk and Moore, 2009).

\subsection{Rhea}

Rhea is the second largest moon of Saturn, with a diameter of $1529$ km and a density of $1.237 gr/cm^3$  (Thomas 2010), a value somewhat greater than that of liquid water. Therefore, it is thought to be composed of a homogeneous mixture of ice and rock. 
Its surface of Rhea is densely cratered with abundant large heavily degraded craters. Rhea seems to be more heavily cratered than Dione and Tethys. An outstanding feature of this moon is a fresh ray crater of $\sim$ 47.2 km of diameter, known as Inktomi. This crater is probably the youngest feature on Rhea's surface, whose estimated age ranges from 8 to 280 Myrs old (Wagner et al., 2008). 
Measurements by Cassini spacecraft have detected a tenuous atmosphere of oxygen and carbon dioxide, which appears to be sustained by chemical decomposition of the surface water ice under irradiation from Saturn's magnetospheric plasma (Teolis et al., 2010).
  Stephan et al. (2012) analyzed  the spectral characteristics of Rhea and concluded that the major process that affects the surface properties of Rhea is the interaction between the surface material and the space environment which includes the impacts of energetic particles from the magnetospheric plasma.  This exogenous process could be responsible for the erosion of small craters on the surface of Rhea.  

\begin{figure}[t!]
\centerline{\includegraphics*[width=0.9\textwidth]{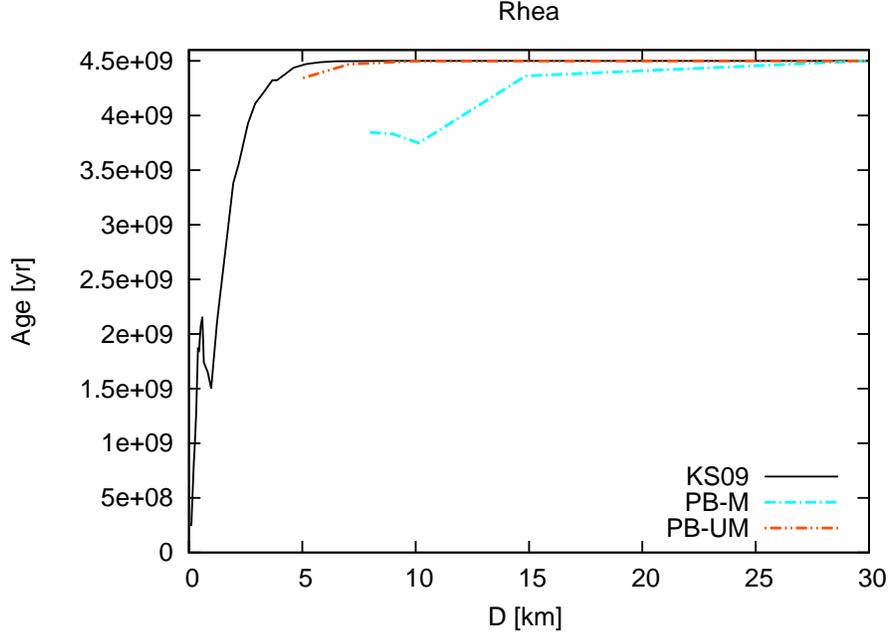}}
\caption{ Age of equatorial mantled (PB-M) and  unmantled regions (PB-UM) from Voyager images and  Cassini images (KS09) of Rhea with respect to D.}
\label{rhea}
\end{figure}

Plescia and Boyce (1982) analyzed  different areas of Rhea's surface and found significantly different  crater density. From our model, we found an age of $4.5$ Gyrs old for the north polar region, this zone being densely cratered. Equatorial regions show  subdued surface texture and absence of small craters, suggesting that some areas have been mantled. Thereby, Plescia and Boyce (1982) separated them into mantled (M) and  unmantled regions (UM).    
KS09 obtained crater counts from Cassini images with  a resolution of  $0.01 - 1$  $km/pixel$ for $\sim 25 \%$ of Rhea's surface, allowing them to reach very small diameters. We calculated ages for this Rhea's surface and plotted them together with the equatorial regions analyzed by Plescia and Boyce (1982) in Fig. \ref{rhea}. We can see that the mantled regions are younger than the other terrains, that is, $\sim 3.8$ Gyrs old for $D \gtrsim 10$ km, consistent with the age estimations by Plescia and Boyce (1985). However, although it is possible that this zone could be younger for smaller sizes, it cannot be guaranteed  since there  are no counts for craters smaller than $8$ km in diameter for these zones. Then, in order to compare Rhea's surface age with the other satellites, we plot our results for KS09 counts in Fig. \ref{todos}.

Rhea has the oldest surface with respect to the other satellites, with the exception of Iapetus (Fig. \ref{todos}). However, craters with  $D \lesssim 4.5 $ km could be young, and the slope of the age curve is steep, which causes age to strongly decrease in direct proportion to crater size. Thus, the possible process that might have erased small craters has a strong dependence on size.  This agrees with 
the above mentioned   process that gradually erodes the surface of Rhea, thus erasing small 
craters without burying the greatest ones. In Fig. \ref{ncno}, for example, there are 
$\sim 40000 $ per $10^6$ km$^2$ craters with  $D \gtrsim 2 $  that are erased from the surface of Rhea in 
the age of the Solar System. 
A peak near $\sim 0.5-$km craters can be seen in the age curve. For $D \lesssim 1$ km, the number of 
observed craters on Rhea  seems to have increased from the theoretical slope (see Fig. 7 in DZ13). Therefore, it is  expected that our calculated age curve will change its behavior. As KS09 noted, for their observed impact crater distributions on the Saturnian satellites,  $D < 1-$km craters are likely to be contaminated with secondary craters at some level, thus this ``peak'' in the age curve is probably due to this contamination on Rhea in the observed crater counts for small diameters.      

In relation to the observed cumulative crater SFD, KS09 and KS10 obtained cumulative slopes  near $ -2$ for all the diameter ranges. Then, although our model with $s_2 = 3.5$, that produces a crater SFD with a power-law index of $\sim -3$, fits better the number of observed craters, the power-law index of the distribution does not.  Therefore, there exists the possibility that the surface of  Rhea, is also saturated for the crater range observed. This would be consistent with the high density of craters observed and with the results of  
 Squyres et al. (1997), who  found that at least $25 \%$ of the craters on Rhea were destroyed by subsequent obliteration.

\subsection{Comparison with other age estimations}

As mentioned before, KS09 obtained the number of craters for the mid-sized icy satellites, and from previous cratering rate estimations by Zahnle  et al. (2003), they calculated surface ages for each satellite and for some crater diameters too. Zahnle  et al. (2003) analyzed two cases of a  population of impactors. In case A,  they inverted crater counts on the Galilean satellites to obtain the size distribution of impactors on Jupiter. In case B, they inverted crater counts on Triton from Voyager images to obtain the size distribution of impactors. In Table \ref{tabla1}, we show both our age estimations for Mimas, Tethys, Dione and Rhea, for some diameters, and the estimations by KS09. The tendency of an increasing  age with diameter that  we have found is only shown by the results of case B (in general). Our results lie, in general, between case A and B, but for small craters they are more similar to case B. This could be due to the fact that the production of craters in the Neptune System could be more directly related to SDOs just entering the Centaur zone, which is the case we are working on. However, the estimations by Zahnle  et al. (2003) have an uncertainty factor of 4.

\begin{table}
\begin{minipage}[t]{\columnwidth}
\caption{ Ages obtained from our results and those calculated by KS09 all in Gyrs for various diameter ranges.}
\label{tabla1}
\centering
\renewcommand{\footnoterule}{}  % to avoi
\begin{tabular}{lrccc} \\
\hline
Satellite & $D $ [km]  &  Age & KS (A)   & KS (B) \\
\hline
\hline
Mimas &  5     & 3.1    &  4.4        & 0.8     \\
        & 10    & 4.2    &  4.4        & 1.3    \\
      &  20    & 4.5    &  3.3        & 0.9     \\
\hline
Tethys &  1    & 1.1    &  4.6   &    1.6      \\
      &   2    & 1.6    &  4.6   &   1.2     \\
       &  5    & 3.5    &  4.6   &    1.7    \\
       &  10   & 4.2    &   4.5  &  2.1  \\
       &  20   & 4.4    &   4.3  &  2.7  \\
\hline
Dione-cp &  1    & 1.2    &  4.6   &    0.9      \\
      &   2    &  2.5    &  4.6   &   1.4     \\
       &  5    &  4.4    &  4.6   &    2.6    \\
       &  10   &  4.5   &   4.6  &  3.3  \\
       &  20   &  4.5    &   4.5  &  3.1  \\
\hline
Dione-sp &  5    &  4.2    &  4.6   &    2.0    \\
       &  10   &  4.5   &   4.4  &  2.0  \\
       &  20   &  4.5    &  3.3  &  1.4  \\
\hline
Rhea &  1      & 1.6     &  4.6   &    1.7      \\
      &   2    &  3.4    &  4.6   &   2.2     \\
       &  5    &  4.5    &  4.6   &    3.1   \\
       &  10   &  4.5    &  4.6   &  3.7  \\
       &  20   &  4.5    &  4.5   &  3.9  \\
\hline
\end{tabular}
\end{minipage}
\end{table}

According to our results, all surfaces analyzed up to now appear be old. However, we have noticed that for small sizes  the observed number of craters is smaller than the calculated number from our model. The size limit of each satellite is presented in Table \ref{dlim}. Although this size limit is different for each satellite, a common explanation for  this paucity of small craters could be crater saturation. In particular, crater saturation starts in small sizes and would gradually incorporate larger crater sizes.

\subsection{Enceladus }
\label{e}

Enceladus is a very interesting and beautiful moon of Saturn. It has a diameter of $504$ km and a density of $1.609 $gr$/$cm$^3$ (Thomas, 2010).  
Cassini-Huygens images reveal a very active surface with unique features, which invites one to investigate them. The discovery of the  plumes of water vapor and small icy particles in the south polar region by Cassini Huygens in 2005  (Porco et al., 2006) showed us a very active world and at the same time, it was very important for the study of icy satellites. High resolution images of the south polar terrains from Cassini reveal tectonic features,  a region almost entirely free of impact craters but with  several  $\sim 130$-km-long fractures called ``tiger stripes'', probably warmed by internal tidally generated heat (Porco et al., 2006). It was also found that this south polar region is 20 K hotter than expected from models, and a thermal emission of 3 to 7 GW was detected (Spencer et al. 2006). 
Gravity field studies by Iess et al. (2014) demonstrated that the structure of Enceladus is compatible with a differentiated body with, for example, a relatively low core density of $\sim 2.4 gr/cm^3$ and an $H_2O$  mantle of density of $ 1 gr/cm^3$ and thickness of 60 km. However, the local south polar heat fluxes, gravity and topography, is more consistent with a regional sub sea of $\sim 10$ km thick in the south pole located beneath an ice crust $\sim 30-40$ km thick (Iess et al., 2014).  
The surface of Enceladus is covered in almost pure water ice (Brown et al., 2006). Newman et al. (2008) mapped the distribution  of crystalline and amorphous ices on the surface of Enceladus from photometric and spectral analysis of data from Cassini Visual and Infrared Mapping Spectrometer. They found a mostly crystalline ice surface with a higher degree of crystallinity in the ``tiger-stripe''cracks, and amorphous ice between these stripes. ``Ice blocks'' were also observed on Enceladus' surface, like the regolites in asteroids and airless bodies of the Solar System (Porco et al., 2006). However, their origin seems to be different. Martens et al. (2015) analyzed their distribution and origin in detail and found that, although they are found in other regions, they are more concentrated within the south polar terrain, where they must be produced mainly by tectonic deformation.   

All this suggests a very active and evolving surface.  Moreover, Enceladus displays different types of terrain: in particular, as one moves away from the ``tiger stripes'', more craters are found (KS09).  This is due to the continuous resurfacing in the south polar zone. In fact, Porco et al. (2006) estimated ages within the south polar terrain possibly as young as 500000 years, or even younger. 

As in the previous section, we estimate the age of Enceladus' surface comparing our calculations of the cratering production with the observed crater counts. There are crater counts on Enceladus from Cassini by KS09, but also previous counts from Voyager by Plescia and Boyce (1983) and Smith et al. (1982).
Images from Voyager spacecrafts 1 and 2 showed a wide diversity of terrains on the surface of Enceladus. Different areas have been identified, according to the populations of craters and other landforms (Fig. 21 of Smith et al., 1982). Among these, the cratered terrains (ct1) and  well preserved terrains (ct2) are distinguished.  Another kind of terrain are the cratered plains (cp), whose principal landforms are bowl-shaped craters. 
In the equatorial region Voyager observed smooth plains sp1 and sp2. The grooves are the principal landforms in sp1, which have few craters of 2 km to 5 km in diameter, while sp2 and ridged plains (rp) are almost free of
craters. The number of observed craters in Voyager images  were taken from  Fig. 1 and Table 1 in Plescia and Boyce (1983) for ct1 and ct2 (ECT) and cp (ECP), and from Fig. 23 in Smith et al. (1982) for sp1, sp2 and rp. 

Kirchoff and Schenk (2009) analyzed Cassini images for crater counts. They analyzed different types of terrain  separately according to variations in crater density and geological features (see their Fig. 4). Those regions are cratered plains (cp-eq and cp-mid) and the ridged regions rp1, rp2, rp3, rp4, rp5, and rp6. 

The regions analyzed by Voyager and Cassini seem to be different. However, through an inspection of cartographic base maps of Enceladus in both papers (Smith et al., 1982 and KS09), it seems that rp and sp2 and probably sp1 regions of Voyager are near the ridged plains (rp1-rp6) of KS09; ct1 and ct2 regions of Voyager are near mid-latitude cratered plains of KS09, and cp regions of Voyager are near equatorial cratered plains of KS09.  
It is worth noting, however,  that the Cassini data have a higher spatial resolution than that from Voyager. The Cassini mission has been able to show craters with $D < 2$ km. Also, KS09 has made a subdivision of ridged plains (rp), resulting in better details. 

\begin{figure}
\centering
\resizebox{\hsize}{!}{\includegraphics{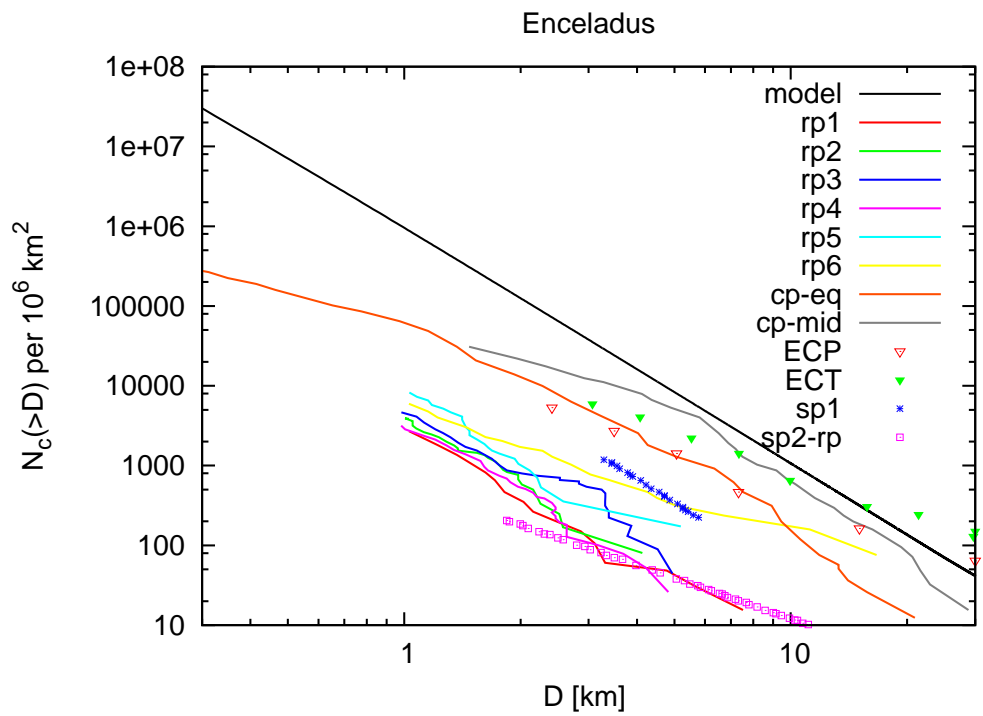}}
\caption{ Cumulative number of craters calculated from model by DZ13 ($s_2 = 3.5$) and crater counts from Voyager (ECP, ECT, sp1, sp2-rp) and from Cassini by KS09. 
%The points represent the two values of the age  calculated by KS09
}
\label{nenc}
\end{figure}

From our model developed in DZ13, we have got the cumulative number of craters produced by current Centaurs. In order to  visualize all the observed crater counts and to compare them with our model, those numbers are shown in Fig. \ref{nenc}.

In general,  the observed curves are below the calculated curve in our model. Therefore, the observed number of craters on the studied terrains would be lower than the current expected contribution of Centaurs with the exception of ECP and ECT, which present a greater number of observed craters than our model for $D \gtrsim 20 - 30$ km respectively.  

\begin{figure}
\centering
\resizebox{\hsize}{!}{\includegraphics{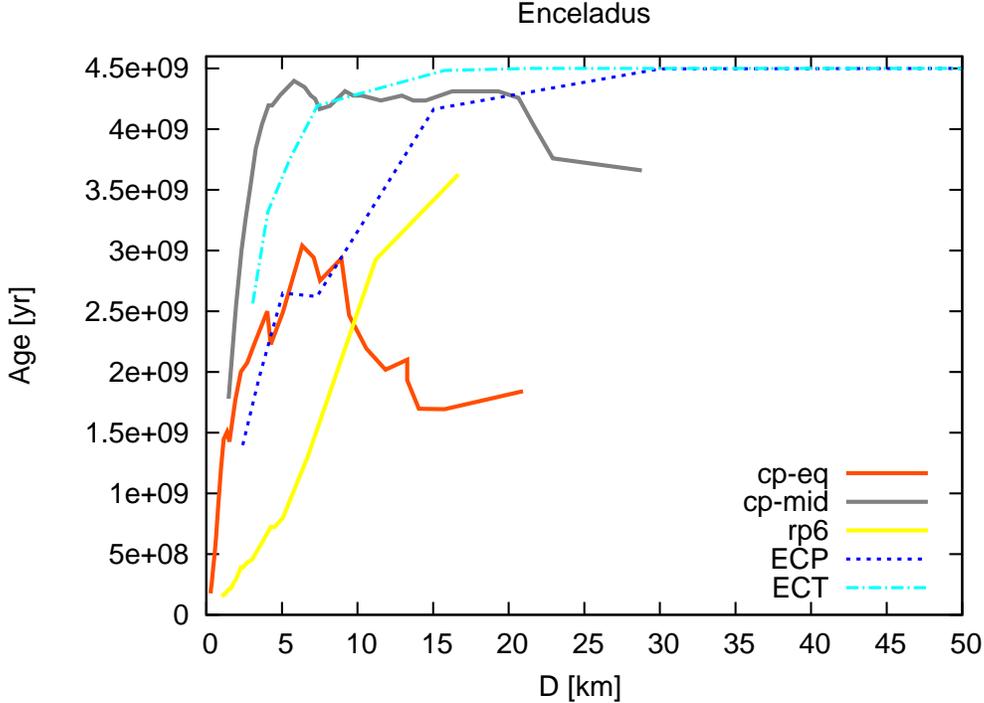}}
\caption{ Age of the cp and rp6  surfaces of Enceladus with respect to D. 
%The points represent the two values of the age  calculated by KS09
}
\label{ageenccp}
\end{figure}

\begin{figure}
\centering
\resizebox{\hsize}{!}{\includegraphics{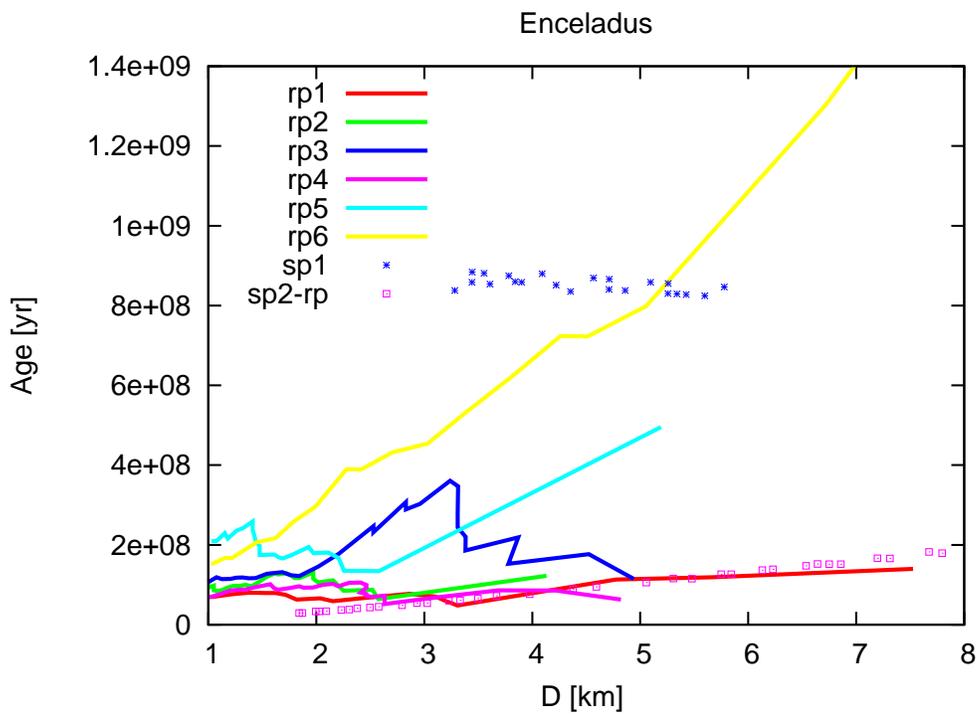}}
\caption{ Age of the rp surfaces of Enceladus  with respect to D. 
%The points represent the two values of the age  calculated by KS09
}
\label{ageencrp}
\end{figure}

We have calculated the age  with respect to D from Eq. \ref{age} and plotted the results in Figs. \ref{ageenccp} and \ref{ageencrp} for both observed crater counts (Voyager and Cassini) and regions. We have separated the results into two figures because of the wide age range. 
We can see that the observed areas of Enceladus are usually young, and that age increases as we move 
away from  the ``tiger stripe'' cracks.  Cratered plains in mid and high latitudes (cp-mid)  are $\sim 4.3$ Gyrs old, regions ct1 and ct2 of Voyager are  $\sim 4.5$ Gyrs old,  but craters less than  $\sim 6$ km could be younger. On the contrary, cratered plains in equatorial latitudes (cp-eq) are younger than 3 Gyrs old for the crater sizes counted by KS09. ECP from Voyager seems to be old, but craters less than $\sim 20$ km could be younger. 

All rp regions appear to be very young. The rp1 region, which is very near the ``tiger stripes'', 
a very active region, is the youngest region, and rp6 is the oldest one. However, all sizes of craters in rp6  are young. In fact, there are craters with $D  \lesssim 16$ km which are younger than $\sim 3.5$ Gyrs old. In rp6, age increases with diameter like in the other mid-sized satellites. But in the rp1-rp5 zones, ages are similar for  all the diameter ranges analyzed and they are all less than $\sim 500$ Myrs old. The very young region rp1 is $\sim 70$ Myrs old on average. Those ages of the ridged plains would be in agreement with estimations by Plescia and Boyce (1985) who estimated an age of no more than 100 Myrs.  This behavior in the rp1-rp5 zones must be due to the plumes emanating from the south polar region and tectonic activity probably associated with ``tiger stripes'', whose production mechanisms strongly erase those craters.

In comparison with previous research, our ages are more similar to those calculated by KS09 for their case B (see Table 3).

Moreover,  Enceladus' plumes are  thought to be the source and maintenance of  E-ring material, but a fraction of the plume particles returns to Enceladus hitting its surface  (Kempf et al., 2010). The particle deposition on Enceladus surface is more frequent in the ``tiger stripes'' than in other regions, as for example mid latitudes  (Kempf et al., 2010). Also, the largest water ice particles covering Enceladus' surface ($\sim 0.2$ mm) are concentrated in the ``tiger stripes'' zone and in general, the particle diameters are strongly correlated with the distance to this zone (Jaumann et al., 2008). Therefore, the plume material returning to Enceladus could be an ``exogenous'' source causing the erasure of craters preferentially near the tiger stripes.

On the other hand, the heat emanating from the south polar terrain, detected by Cassini's Composite Infrared Spectrometer, suggests internal geological activity (Spencer et al., 2006). The heat source for the production of the plumes, resurfacing and heat flow in the south polar region of Enceladus, has not yet been sufficiently understood. The estimated radiogenic and tidal heating resulting from the orbital eccentricity of Enceladus are not sufficient to account for the energy emanating from the south polar zone (Porco et al., 2006; Meyer and Wisdom, 2007). Besides, the absence of internal activity on Mimas, which has an orbital eccentricity greater than Enceladus', further constrain this as an unlikely source. Porco et al. (2006) analyzed other tidal heating mechanisms, such as mean motion resonances among the Saturnian moons, the 2:1 mean motion resonance with Dione being the most important, and secondary resonance heating, but those mechanisms are also inadequate. Squyres et al. (1983) suggested that the current rate of tidal heating might be sufficient to maintain a liquid sea under a crust of 10 km, but 
a previous stronger heating would be required to initiate the melting. This previous process  might have been that Enceladus had a larger eccentricity in the past (to obtain a greater tidal heating) or was captured into mean motion or secondary resonances with other satellites (Porco et al., 2006; Meyer and Wisdom, 2008).  Meyer and Wisdom  (2008) demonstrated that Enceladus is probably near its equilibrium eccentricity and that the current equilibrium heating rate is not enough to trigger the melting. Therefore, any previous stronger heating requires some non-equilibrium behavior.

\section{Discussion and conclusions}

Based on both calculations of the cratering rate and the number of craters produced by Centaurs, and the comparison with observations, we have estimated the surface age of each observed terrain on each mid-sized satellite of Saturn. 
The results are given for each satellite, but in general terms we can say that there are less observed small craters than calculated (except on Iapetus), which result in young ages calculated from our model. This could be interpreted as either efficient endogenous or exogenous  process(es) erasing small craters or crater saturation. 

We have also found that the smaller craters are being preferentially removed.  This agrees with a gradual process of erosion, either from a geological or exogenous origin  or by cratering itself. 
This erasing process does not affect large craters, and the size limit depends on the satellite. For craters greater than this size limit, our results imply that there are  primordial craters, and/or that another main source of craters might exist.

We have also found a correlation between the limit size  from which small craters are erased (and then age is below 4.5 Gyrs old)  and the distance to Saturn. That is, 
 the smaller the distance from Saturn, the higher the size limit (see Table \ref{dlim}).  It is possible that there is not a general process to describe this behavior since each satellite has its own peculiarities. However, a potential common process that seems to explain the paucity of small craters  is crater saturation which could be connected to the correlation between distances to Saturn and erasure size limit. Another possibility is that deposition of  E-ring  particles could be responsible for such a correlation. Since the E-ring has a density that generally decreases from Enceladus to Rhea  (Hor\'anyi et al., 2008), the E-ring flux on the mid-sized satellites could be lower as  we move away from Enceladus.

The estimated ages for the terrains analyzed on the mid-sized satellites, with the exception of Enceladus, are  $\sim 4.5$ Gyrs old. Our ages are, within the ages calculated by KS09, but the fact that smaller craters are  preferentially removed seems to be strong in our calculations.  Almost all the Enceladus' terrains could be  young, with the exception of the mid and high latitude plains. The regions near the south polar terrain could be as young as 50 Myrs old.
On the contrary, all the surface of Iapetus is old and it likely records a primordial  source of craters, smaller craters being  likely approaching  saturation until  $D \lesssim 2$ km-craters, where saturation is complete. 

The ages calculated here are based on the current number of Centaur objects as contributors to craters. We have used the best known estimations from the literature for both this number and the size-frequency distribution. However, for small sizes, this SFD is uncertain and also unreachable with current direct observations. That is why the theoretical crater contribution, the comparison with observations and the determination of terrain ages in this paper should be carefully interpreted  and not taken as absolute ages.
They could serve, instead, to enhance both the understanding of the processes occurring on the satellites and the SFD for small sizes. Finally, the comparison of our age estimations with geological time scales on Saturnian satellites could help  us determine what is still unknown today about those wonderful worlds.\\

\noindent{\bf Acknowledgments:} We would like to thank  Gabinete de Ingl\'es de la Facultad de Ciencias Astron\'omicas y Geof\'{\i}sicas de la UNLP for  a careful language revision. We are grateful to  Eduardo Fern\'andez Laj\'us  for valuable discussion suggestions and to Michelle Kirchoff and an anonymous referee for valuable comments and suggestions that helped us improve the manuscript.

\end{document}